\documentclass[aps,prb,twocolumn,floatfix,superscriptaddress,showpacs,footinbib]{revtex4}
\usepackage{amsmath,amssymb,natbib,bm,graphicx,url,epsfig,epstopdf,bbm}
\usepackage{ulem}
\usepackage[ansinew]{inputenc}

\usepackage{color}

\begin{document}

\title{Effect of a Coulombic dot-lead coupling on the dynamics of a quantum dot} 

\author{Florian Elste}
\affiliation{Department of Physics, Columbia University, 538 West 120th Street, New York, NY 10027, USA}

\author{David R. Reichman}
\affiliation{Department of Chemistry, Columbia University, 3000 Broadway, New York, NY 10027, USA}

\author{Andrew J. Millis}
\affiliation{Department of Physics, Columbia University, 538 West 120th Street, New York, NY 10027, USA}

\date{\today}

\begin{abstract}
The effect of a Coulombic coupling on the dynamics of a quantum dot hybridized to leads is determined.  The calculation treats the interaction between charge fluctuations on  the dot and the dynamically generated image charge in the leads.  A formally exact solution is presented for a dot coupled to a Luttinger liquid and an approximate solution, equivalent to treating the lead dynamics within a random phase approximation, is given for  a dot coupled to a two- or three-dimensional metallic lead.  The leading divergences arising from the long-ranged Coulomb interaction are found to cancel, so that in the two- and 
three-dimensional cases the quantum-dot dynamics is equivalent to that obtained by neglecting both the dot-lead Coulomb coupling and the Coulomb renormalization of the lead electrons, while in the one-dimensional case the dot-lead mixing is enhanced relative to the noninteracting case.
Explicit results are given for the short-time dynamics.

\end{abstract}


\maketitle

\section{Introduction}

The quantum dot, a system comprised of a small number of spatially localized levels coupled to one or more metallic leads, is one of the paradigmatic problems of condensed matter theory  and is highly relevant to nanoscience.  Theoretical studies of quantum dots typically involve two competing effects: local interactions, which constrain the possible electronic configurations of the quantum dot, and hybridization with the leads, which mixes different dot eigenstates. In this paper we study a third crucial and physically relevant interaction: the Coulomb coupling between the charge on the dot and the charge on the leads.  This interaction is most important in the case of a dot weakly coupled to leads, because in this case the physical situation is of infrequent transitions between states of well defined integer charge. 

Short-ranged dot-lead interactions play a crucial role in X-ray absorption lineshapes  \cite{Mahan67} and the Kondo effect\cite{Anderson70} and were studied using bosonization methods by Schotte and Schotte.\cite{Schotte69} But the general issue of dot-lead interactions received relatively little attention in the recent nanoscience literature, although Gefen and co-workers have noted that introducing an additional Coloumbically coupled lead into a standard quantum-dot problem may lead to a non-Fermi-liquid state.\cite{Gornyi,Gutman,Gutman2,Dinh}

The interaction we wish to treat may be written as
\begin{equation}
H_\text{Coul}= n_d \sum_{a}\int d^D r {\rho}_a(r) \frac{e^2}{\epsilon r}.
\label{Hcoul}
\end{equation}
Here we denote by  $n_d$ the operator giving the number of electrons on the dot and by $\rho_a(r)$ the operator giving the number density of electrons at position $r$ relative to the impurity in channel $a$  of a $D$-dimensional lead. $\epsilon$ is a background dielectric constant. Consistency requires that a theory involving  $H_\text{Coul}$  involves also considerations of the Coulomb effects on the density-density response of the lead electrons.  In the case of a one-dimensional lead, the standard techniques of bosonization\cite{Haldane81} allow us to include both effects,  expressing the low-energy electronic physics entirely in terms of density and spin fluctuation operators and enabling a complete and formally exact theory.  We find that the interaction has (as expected from previous results obtained for the X-ray edge problem \cite{Mahan67,Schotte69}) a profound effect on the dynamics of the impurity, qualitatively changing the power laws describing the time evolution.  We also present an extension to the case of higher-dimensional leads. 
We employ a canonical transformation first introduced by Bohm and Pines,\cite{Bohm53} which can only be carried out approximately but which  captures the essential physics of screening.  In this case we find that the Coulombic renormalizations of the dot-lead coupling and the lead-electron dynamics cancel, so the dot-lead problem can be treated within an essentially noninteracting electron approximation.

The physics of local defects in Coulomb Luttinger \mbox{liquids} has received previous attention.
Fabrizio~\textit{et al.}\cite{Fabrizio}~and Maurey and Giamarchi\cite{Maurey}
have considered the effects of one or more impurities (described as a short-ranged
potential scatterer) on a Luttinger liquid with Coulomb interaction.
Liu\cite{Liu} has investigated the screening of a test charge, however, in
a model with only short-ranged electron-electron interactions.
None of these authors treated an impurity with dynamical charge fluctuations or a
Coulombic dot-lead coupling. We also note that Lerner~\textit{et al.}\cite{Lerner}
have studied a model related to the specific model we consider, namely, an
impurity adjacent to a Luttinger liquid. They did not consider the Coulombic coupling
but did include charge fluctuations and found a nontrivial structure of the
transmission coefficient.

The effect of a dot-lead interaction on the population of a quantum dot has been studied by Goldstein \textit{et al.}\cite{Goldstein} using density matrix renormalization group and classical Monte Carlo simulations. Particular attention was paid to the screened case where the long-range part of the Coulomb interaction can be neglected. The electronic tunneling was investigated using a Coulomb-gas analysis, which allows for an expansion to all orders in the dot-lead hybridization. The enhancement of the electronic tunneling at the resonance found in Ref.~\onlinecite{Goldstein} has a similar physical interpretation as the results presented in this paper.
 
The paper is organized as follows. In Sec.~\ref{Model} we introduce the model. Section~\ref{OneD} presents the analysis of the  one-dimensional problem, including the  explicit forms appropriate to a dot with Luttinger liquid leads. Section~\ref{3D} presents the  extension of the results to the case of two- and three-dimensional leads.  An application of the results to the short-time impurity dynamics is presented in  
Sec.~\ref{Dynamics}. We summarize our results in Sec.~\ref{Conclusions}.

\section{Model \label{Model}}

We consider a quantum dot that is hybridized to one or more leads, but neglect dot-lead potential scattering  and do not consider the case where the dot breaks the lead into two semi-infinite leads connected only by hybridization through the dot. These simplifications allow us to  focus on  the consequences of the Coulombic coupling $H_\text{Coul}$, Eq.~(\ref{Hcoul}). The effects we have neglected introduce additional 
complications \cite{Fabrizio,Maurey,Liu,Lerner,Kane92} whose interplay with the Coulombic coupling will be treated in a future paper. 

The quantum-dot problem is described by a Hamiltonian of the general form 
\begin{equation}
H = H_\text{lead} + H_\text{dot} +  H_\text{Coul}+H_\text{mix}
\label{H}
\end{equation}
with $H_\text{Coul}$ given by Eq.~(\ref{Hcoul}). The quantum-dot Hamiltonian $H_\text{dot}$ may be written as 
\begin{equation}
H_\text{dot}= \varepsilon_d \, n_d +\frac{U}{2} n_d(n_d-1) +\dots
\label{Hdot}
\end{equation}
Here $U$ is the dot charging energy and $n_d=\sum_{\alpha\sigma} d_{\alpha \sigma}^\dagger d_{\alpha \sigma}$ is the operator giving the total number of electrons on the dot. The operator $d_{\alpha \sigma}^\dagger$ creates an electron with energy $\varepsilon_d$ and spin $\sigma$ in dot state $\alpha$. The ellipsis denotes other on-dot interactions, for example, the Hund's coupling $J$.

We label the lead orbitals by $a$ and write the lead Hamiltonian as
\begin{align}
H_\text{lead} &~=~ \sum_{a k \sigma} \epsilon^a_{k} c^{\dagger}_{a k \sigma} c_{a k \sigma}
\nonumber \\
&~+ \frac{1}{2}\int d^Dr \, d^Dr' \frac{e^2}{\epsilon|r-r'|} {\rho}(r){\rho}(r')+\dots,
\label{Hlead}
\end{align}
where $c^{\dagger}_{a k \sigma}$ creates an electron with momentum $k$ and spin $\sigma$ in state $a$, $\rho(r)$ is the  operator giving the charge density at position $r$, $\epsilon$ is a background dielectric constant and the ellipsis denotes any additional short-ranged interactions.  It is important that the same long-ranged interaction appears in Eqs.~(\ref{Hcoul}) and (\ref{Hlead}).

The dot-lead hybridization is given by 
\begin{equation}
H_\text{mix} = \sum_{a k \sigma \alpha} \left[ \mathcal{T}^{a k \sigma}_\alpha \, d^{\dagger}_{\alpha \sigma} c_{a k \sigma} + {\mathcal{T}^{a k \sigma}_\alpha}^* \, c^{\dagger}_{a k \sigma} d_{\alpha \sigma} \right].
\end{equation}

The standard quantum-dot physics arises because  $\left[H_\text{mix},H_\text{dot}\right]\neq 0$ so that the interaction-induced constraints on the dot occupancy interact nontrivially with the hybridization, giving rise, for example, to the Kondo effect. The physics we wish to investigate arises because $\left[H_\text{mix},H_\text{Coul}\right]\neq 0$ so that a hybridization event changes the local charge, giving rise to a long-ranged electric field to which the lead electrons react.  

\section{One dimension \label{OneD}}

If the leads are one-dimensional and the electron dispersion is linear, then the low-energy physics of $H_\text{lead}$ may be expressed in terms of bosons \cite{Luther,Haldane81} and the results used to solve the model exactly. We illustrate the method here for a single lead with multiple channels; the generalization to multiple leads is straightforward but involves more complicated algebra.

We imagine a system with linear dimension $L$ (which we will later take to infinity) and periodic boundary conditions so the allowed values of $q$ are $2\pi n/L$ with $n\neq 0$ an integer. We combine spin and orbital quantum numbers into a superindex $\beta=1,\dots,M$. The physics is conveniently represented in terms of right ($\lambda=+$) and left ($\lambda=-$) moving particle-hole pairs
$\rho^\pm_{\beta}(q)$,
which obey the commutation relation 
$[\rho^\pm_{\beta}(q),\rho^\pm_{\beta}(-q')]=\pm \delta_{qq'} qL/2\pi$.\cite{Giamarchi}
These can be recombined into boson operators, 
\begin{align}
{\phi}_\beta(q) &= -\frac{i}{q} \, \left[ \rho^+_\beta(q) + \rho^-_\beta(q)\right], \label{Eq19} \\ 
\Pi_\beta(q) &= - \left[ \rho^+_\beta(q) - \rho^-_\beta(q) \right], \label{Eq21} 
\end{align}
which obey the volume commutation relation
\begin{equation}
\left[\phi_\beta(q),\Pi_{\beta'}(-q')\right]=i\frac{L}{\pi} \, \delta_{\beta\beta{'}}
\delta_{qq'}.
\label{commutation}
\end{equation} 
The total particle density in lead $\beta$ is given by
\begin{equation}
\rho_\beta(q)=iq\phi_\beta(q).
\end{equation}

The lead electron creation operator $\psi_{\lambda\beta}$ may also be expressed in terms of bosons as
\begin{equation}
\psi_{\lambda \beta}(x)=\frac{1}{\sqrt{2\pi\eta}}e^{i\lambda k_F x}e^{ i \frac{\pi}{L}\sum_q e^{iqx} \left[ \lambda{\phi}_{\beta}(q)-\frac{1}{iq} \Pi_{\beta}(q)
\right]}.
\label{psidef}
\end{equation}
Here we have omitted the Klein factors that carry the Fermi statistics and have introduced a small positive infinitesimal factor $\eta$ arising from the correct normal ordering of the operators.\cite{Haldane81}

A key result of the theory of one-dimensional conductors is that in the absence of Umklapp scattering  the low-energy physics of the leads  may be described by new boson operators $\phi_b,\Pi_b$ related by a linear transformation to the operators $\phi_\beta,\Pi_\beta$ and  also obeying the canonical commutation relations Eq.~(\ref{commutation}). In terms of the new operators the lead Hamiltonian becomes 
\begin{align}\label{Hamiltonian3}
H_\text{lead} = \sum_{b=1,\dots,M} \sum_q \frac{\pi}{2 L} v_b(q) & \bigg[ K_b(q) \, \Pi_b(-q) \Pi_b(q) \nonumber \\
& + \frac{q^2}{K_b(q)} \phi_b(-q) \phi_b(q) \bigg]
\end{align}
with velocity  parameters $v_b$ and interaction parameters $K_b$ determined by the bare velocities and interactions of the lead eigenstates.

In the most general case the transformation relating the $\phi_b,\Pi_b$ to the $\phi_\beta,\Pi_\beta$ is complicated; in particular the equations for $\phi_b$ and $\Pi_b$ each may involve both $\phi_\beta$ and $\Pi_\beta$ and the final $b$ combinations need not have a simple interpretation in terms of the original densities $\psi^\dagger_\beta\psi_\beta$. However in the most relevant case, where all of the channels in a given lead have the same bare velocity and the interactions conserve the total lead density, then one of the channels (which we take to be $b=1$ for definiteness) is (up to an overall factor) the total charge density and is  given by
\begin{eqnarray}
\phi_{b=1}(q)&=&\frac{1}{\sqrt{M}}\sum_{\beta}\phi_\beta(q),
\\
\Pi_{b=1}(q)&=&\frac{1}{\sqrt{M}}\sum_{\beta}\Pi_\beta(q).
\end{eqnarray}
Conversely, for any of the original indices  $\beta$ we have 
\begin{eqnarray}
\phi_{\beta}(q)&=&\frac{1}{\sqrt{M}}\phi_{b=1}(q)+\dots,
\\
\Pi_{\beta}(q)&=&\frac{1}{\sqrt{M}}\Pi_{b=1}(q)+\dots,
\end{eqnarray}
with the ellipses representing the other operators (all commuting with $\phi_{b=1},\Pi_{b=1}$) needed to make up the full operator.  In particular, the electron annihilation operator assumes the form  \cite{Haldane81}
\begin{align}
\psi_{\lambda\beta}(x)& ~=~ e^{i\frac{\pi}{\sqrt{M}L} \sum_q e^{iqx} \left[ \lambda{\phi}_{1}(q)-\frac{1}{iq} \Pi_{b=1}(q)\right]} \, \psi_{\lambda\beta}^\text{rest}(x)
\label{psidef1}
\end{align}
with
\begin{align}
\psi_{\lambda\beta}^\text{rest}(x) &~=~ \frac{e^{i\lambda k_F x}}{\sqrt{2\pi\eta}} \,   \nonumber \\
&\times e^{i \frac{\pi}{\sqrt{M}L} \sum_{b=2}^{M}\sum_q e^{iqx} \left[ \lambda C_{\beta b}{\phi}_{b}(q)-\frac{1}{iq} D_{\beta b}\Pi_b(q)\right]}.
\end{align}
with $C$ and $D$ the transformation coefficients which diagonalize the Luttinger-liquid Hamiltonian.

Comparison to Eq.~(\ref{Eq19}) shows that the total charge density $\rho$ is related to $\phi_{b=1}$ by 
\begin{equation}
\rho(q)=\sqrt{M}iq\phi_1(q).
\label{rhoofphi}
\end{equation}
Thus writing
\begin{eqnarray}
\frac{1}{|x|}&=&\frac{1}{L}\sum_qe^{iqx} W_q, \\
W_q&=& \ln\left(1+\frac{\Lambda^2}{q^2}\right)
\label{Wdef}
\end{eqnarray}
with $\Lambda$ the inverse of a short-distance cutoff we find that the long-ranged Coulomb interaction between conduction electrons in the Luttinger liquid is
\begin{equation}
H_\text{int}=\frac{M \pi v_F V_c }{2L} \sum_q q^2\phi_1(q)\phi_1(-q) W_q.
\label{Hint1d}
\end{equation}
Here we introduced a dimensionless measure of the Coulomb interaction strength 
\begin{equation}
V_c=\frac{e^2}{\pi v_F\epsilon}.
\label{Vcdef}
\end{equation}
In a general Coulomb-coupled Luttinger liquid we have also a short-ranged part of the interaction, parametrized by dimensionless constants $g_{1,2}$ such that if the Coulomb interaction were negligible we would have
\begin{eqnarray}
K_{1,SR}(q)&=&\sqrt{\frac{1+g_1-g_2}{1+g_1+g_2}},
\label{K1SRdef}\\
v_{1,SR}(q)&=&v_F K_{1,SR}(q) \left(1+g_1+g_2\right).
\label{v1SRdef}
\end{eqnarray}
Including the Coulomb interaction gives 
\begin{eqnarray}
K_1(q)&=&\frac{K_{1,SR}(q)}{\sqrt{1+\frac{MV_cW_q}{1+g_1+g_2}}},
\label{K1def}\\
v_1(q)&=&v_{1,SR}(q)\sqrt{1+\frac{MV_cW_q}{1+g_1+g_2}}.
\label{v1def}
\end{eqnarray}

The dot-lead interaction, Eq.~(\ref{Hcoul}), is transcribed into the new representation as 
\begin{equation}
H_\text{Coul}=\frac{\pi v_F}{\sqrt{M}L}\sum_qiq\phi_1(q) \, MV_cW_q \, n_d.
\label{Hcoul1d}
\end{equation}
The linear coupling between the dot occupancy $n_d$ and the lead density $\phi_1$ in 
Eq.~(\ref{Hcoul1d}) may be removed by a canonical transformation which shifts $\phi_1(q)\rightarrow \phi_1(q)-\frac{n_d}{iq\sqrt{M}} Z_q$ with 
\begin{equation}
Z_q=\frac{v_FK_1(q)}{v_1(q)}MV_cW_q=\frac{MV_cW_q}{1+g_1+g_2+MV_cW_q}.
\label{Zdef}
\end{equation}

In the shifted Hamiltonian the dot-lead interaction is eliminated,  $H_\text{lead}$ retains the form of Eq.~(\ref{Hamiltonian3}) and the dot energy $\varepsilon_d$ and the local interaction $U$ are decreased by $\Delta$ and $2\Delta$ respectively, with the \textit{polaron shift} $\Delta$ given by
\begin{eqnarray}
\Delta&=&\frac{v_F}{2M}\frac{\pi}{L}\sum_q\frac{v_1(q)}{v_FK_1(q)}Z^2_q
\nonumber \\
&=&\frac{v_F}{2M}\frac{\pi}{L}\sum_q\frac{(MV_cW_q)^2}{1+g_1+g_2+MV_cW_q}.
\label{deltadef}
\end{eqnarray}

Here the polaron shift gives the static interaction between the dot charge and the image charge it induces in the lead.

The canonical transformation acts on an operator ${\cal O}$ by ${\cal O}\rightarrow e^{iS}{\cal O}e^{-iS}$. From Eq.~(\ref{commutation}) we see that 
\begin{equation}
S= - n_d \frac{\pi}{\sqrt{M}L}\sum_q\Pi_1(-q)\frac{Z_q}{iq}.
\label{Sdef}
\end{equation}
Under the canonical transformation the fermion operator $\psi_{\lambda\beta}(x)$, Eq.~(\ref{psidef1}), becomes
\begin{equation}
\psi_{\lambda\beta}(x)\rightarrow e^{-i\lambda  n_d \frac{\pi}{M L}\sum_q \frac{e^{iqx}}{iq}Z_q}\ \psi_{\lambda\beta}(x)
\end{equation}
while  the operator $d_{\alpha \sigma_\beta}^\dagger$ becomes
\begin{equation}
d^\dagger_{\alpha \sigma_\beta}\rightarrow d^\dagger_{\alpha \sigma_\beta} e^{- i \frac{\pi}{\sqrt{M}L}\sum_q \Pi_1(-q)\frac{Z_q}{iq}}.
\end{equation}
The factor multiplying $\psi$ is purely imaginary. In this paper we will only need to consider combinations $d^\dagger \psi \psi^\dagger d$  for which this factor and a similar one coming from the commutator needed to combine the factor multiplying $d$ with the boson operators in $\psi$ cancel. We may thus write the transformed operator 
appearing in the dot-lead hybridization term as
\begin{align}
e^{iS} \, d^\dagger_{\alpha \sigma_\beta}  & \psi_{\lambda\beta}(x) \, e^{-iS}~=~ d^\dagger_{\alpha \sigma_\beta} \ \psi_{\lambda\beta}^\text{rest}(x)
\nonumber \\
& \times  e^{i \frac{\pi}{\sqrt{M}L} \sum_q  \left[e^{iqx} \lambda{\phi}_{1}(q)-\frac{e^{iqx}-Z_q}{iq}\Pi_{1}(q)\right]}.
\label{transformedV}
\end{align}
Thus in a dot Coulombically coupled to a one-dimensional conductor the explicit  Coulombic dot-lead coupling may be eliminated by a canonical transformation. The physics associated with the Coulomb interaction is expressed via a renormalization of 
the dot-lead hybridization.

As will be seen below, for the evaluation of physical quantities the crucial objects are lead-operator expectation values of the form
\begin{equation}
\label{Fdef}
F(x,t)=\left<\xi_{\lambda\beta}(x,t) \xi^\dagger_{\lambda\beta}(0,0)\right>,
\end{equation}
where $\xi_{\lambda\beta}$ is the right-hand side of Eq.~(\ref{transformedV}) with the $d^\dagger$ operator removed.

For later use we present explicit formulae for the two most interesting cases: a metallic nanotube lead and a single-channel Luttinger liquid with $\text{SU(2)}$ spin rotation invariance, specializing further  to the case of a   local dot-lead hybridization ($x=0$). In both of these cases the interactions and velocities  for the noncharge channels may be approximated by the free-fermion values of $v_b=v_F$ and $K_b=1$. It is convenient to multiply and divide by the free-fermion correlator $F_0$, which is proportional to $1/t$ at long times and has an appropriate short-time cutoff. We  obtain
\begin{equation}
F(t)=F_0(t)e^{\Phi(t)},
\label{fdef}
\end{equation}
where
\begin{equation}
\Phi(t)=\frac{2\pi}{ML}\sum_{q\neq 0}\frac{1-e^{-i\omega_q^0 t} - B_q\left(1-e^{-i\omega_qt}\right)}{2|q|}
\label{Phidef}
\end{equation}
and
\begin{eqnarray}
\omega_q&=&v_1(q)|q|, \hspace{0.4in} \omega^0_q=v_F|q|,
\\
B_q&=&\frac{K_1(q)+\frac{1}{K_1(q)}\left(1-Z_q\right)^2}{2}.
\label{Bdef}
\end{eqnarray}
In these formulae the Luttinger-liquid correlations are expressed by the factors $K_1(q)$ and $\omega_q/\omega^0_q$ while the effect  of the dot-lead interaction is carried by $Z_q$.

\begin{figure}[!t]
\begin{center}
\includegraphics[width=7.8cm,angle=0]{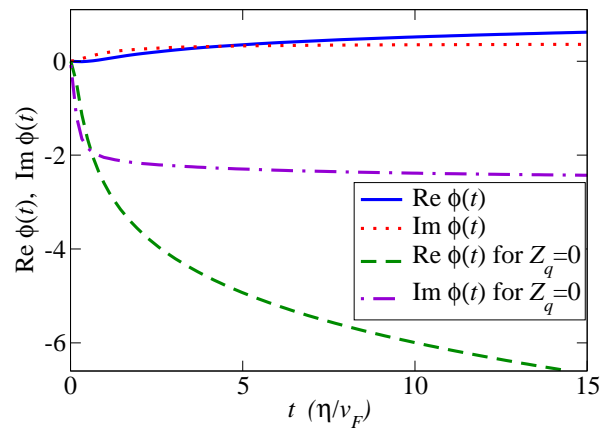} 
\caption{(Color online) Real and imaginary part of the renormalization factor $\Phi(t)$, Eq.~(\ref{Phidef}), for a Luttinger liquid with $M=4$ channels and Coulomb cutoff (tube diameter) $\Lambda=10/\eta$ plotted against time (in units of bare cutoff $\eta/v_F$). Solid blue line: $\text{Re}\,\Phi(t)$ computed from exact expression with dimensionless coupling $V_c \simeq 0.9$. 
Dashed green line: $\text{Re}\,\Phi(t)$ computed from Eq.~(\ref{Phidef}) but with screening factor $Z_q$ set to zero.
Dotted red line: $\text{Im}\,\Phi(t)$ computed from exact expression with $V_c \simeq 0.9$. 
Dash-dotted purple line: $\text{Im}\,\Phi(t)$ computed with screening factor $Z_q$ set to zero.}
\label{phi}
\end{center}
\end{figure}

The real and imaginary parts of $\Phi$ are plotted in Fig.~\ref{phi} for parameters appropriate to a nanotube (see below). In the absence of the Coulombic dot-lead interaction ($Z_q=0$), one finds $\text{Re}\, \Phi(t)<0$ reflecting the suppression of the 
electronic tunneling
by  interaction effects in one dimension.   However,  Eq.~(\ref{Bdef}) shows that the Coulombic dot-lead interaction  acts to reduce the magnitude of the negative term in $\Phi$. We see from Fig.~\ref{phi} that the dot-lead coupling  in fact changes the sign of $\text{Re} \, \Phi(t)$, therefore enhancing the dot-lead hybridization above the noninteracting value. The long-time asymptotic behavior of $\text{Re}\, \phi(t)$ is $\log ( v_F t /\eta)/M$ in the screened case ($Z_q\neq 0$).  

A metallic nanotube has two conducting channels and two spin 
states, so $M=4$. For a metallic nanotube with $\epsilon=1$ and bare Fermi velocity $v_F=5.3 \, \mathrm{eV}\, \mathrm{\AA}$, one has $V_c\approx  0.9$. We also remark that the parameter $\Lambda$ in Eq.~(\ref{Wdef}) is on the order of the inverse of the tube diameter, which is much greater than the basic lattice constant. Further, in metallic nanotubes the short-ranged interactions (and indeed all interaction effects except the long-ranged Coulomb interaction) are believed to be negligible\cite{Kane97} so that $K_{c,SR}=1$ and $v_{c,SR}=v_F$ and 
\begin{eqnarray}
K_1^\text{nanotube}(q)&=&\sqrt{\frac{1}{1+MV_cW_q}},
\label{K1nanodef}\\
v_1^\text{nanotube}(q)&=&v_F\sqrt{1+MV_cW_q}.
\label{v1nanodef}
\end{eqnarray}

Finally, evaluating the polaron shift we find  (choosing the cutoff $\Lambda$ to be the inverse of the nanotube diameter $d_\text{nanotube}$)
\begin{equation}
\Delta_\text{nanotube}[\text{eV}]\approx \frac{5.6}{d_\text{nanotube} [\text{\AA}]}.
\end{equation}
For a typical nanotube diameter of $12\,\text{\AA}$, one obtains $\Delta_\text{nanotube} \sim 0.5\,\text{eV}$.

\section{Lead dimension greater than 1 \label{3D}}

The treatment of the previous section relied on a particular feature of one-dimensional (non-nested) systems, namely that the electronic degrees of freedom could be entirely eliminated in favor of a set of effectively noninteracting  boson excitations, one of which is  the density.  In dimension higher than one a complete elimination of electronic degrees of freedom  is not possible but a  separation of density fluctuation and fermionic variables may be effected via  a canonical transformation method introduced by Bohm and Pines.\cite{Bohm53} The canonical transformation cannot be carried out exactly but an approximate implementation can be performed, which has the same level of accuracy as the familiar random phase approximation (RPA). This approximate implementation leads to a theory which is very similar in form to the one-dimensional theory derived above, but of course expressing the characteristic physics of higher-dimensional systems.

The analysis proceeds from the Hamiltonian in Eq.~(\ref{H}). Following Bohm and Pines\cite{Bohm53} we introduce canonically conjugate boson fields $P_\textbf{q}$ and $Q_\textbf{q}$ obeying (in $D$ spatial dimensions)
\begin{equation}
 \left[ Q_{\textbf{q}}, P_{\textbf{q'}} \right] = i \frac{L^D}{\pi^D}\delta_{\textbf{q}\textbf{q}'}.
 \label{commutator}
 \end{equation}
To begin we consider states $|\psi \rangle $ which are a direct product of fermion and boson eigenstates and restrict attention to states satisfying
\begin{equation}
P_{\textbf{q}}|\psi \rangle = 0.
\label{constraint}
\end{equation}
We then may shift the density operators of all lead states $a$, 
\begin{equation}
\rho_a(\textbf{q}) \rightarrow \rho^\text{shift}_a(\textbf{q})\equiv\rho_a(\textbf{q}) + \frac{P_{\textbf{q}}}{N\sqrt{V_\textbf{q}}},
\end{equation}
where $N$ denotes the number of lead states and $V_\textbf{q}$ denotes the Fourier transform of the Coulomb interaction $4\pi e^2/(q^2L^3)$ in $D=3$ and $2e^2\pi/(|q|L^2)$ in $D=2$ dimensions. Provided that we consider only wave functions $|\psi \rangle$ which obey the subsidiary condition Eq.~(\ref{constraint}) the Hamiltonian written in terms of $\rho^\text{shift}$ is equivalent to the original Hamiltonian.  

Bohm and Pines now introduce a canonical transformation to shift $P_\textbf{q}$ by $-\sqrt{V_\textbf{q}}\sum_a\rho_a(\textbf{q})$.  This transformation is effected by  $H\rightarrow e^{iS}He^{-iS}$ with 
\begin{equation}
S=\frac{\pi^D}{L^D}\sum_{a\textbf{q}} \sqrt{V_\textbf{q}}Q_\textbf{q} \rho_a(\textbf{q}).
\label{S3d}
\end{equation}
After this transformation, the  subsidiary condition for the wavefunctions becomes $[ P_\textbf{q} - \sqrt{V_\textbf{q}} \sum_a \rho_a(\textbf{q})] |\psi \rangle = 0$, which allows us to replace $\sum_a \rho_a(\textbf{q})$ by $P_\textbf{q}/\sqrt{V_\textbf{q}}$. 

Under the transformation, $H_\text{dot}$ remains invariant while $H_\text{mix}$ becomes
\begin{equation}
H_\text{mix}\rightarrow\sum_{a\textbf{k} \sigma} \left[ \mathcal{T} \, e^{-i\frac{\pi^D}{L^D}\
\sum_{\textbf{q}} \sqrt{V_\textbf{q}} Q_{\textbf{q}}} d^{\dagger}_\sigma c_{a \textbf{k} \sigma} + \text{h.c.} \right]
\label{Hmix3d}
\end{equation}
and the excitonic dot-lead coupling $H_\text{Coul}$ is
\begin{equation}
H_\text{Coul} = \sum_{\textbf{q}} \sqrt{V_\textbf{q}} P_{\textbf{q}} n_d.
\label{Hcoul1}
\end{equation}
Similarly, the lead-electron Green function $G_\text{lead}(\textbf{r},t) = \langle c_{a\sigma}(\textbf{r},t) c_{a\sigma}^\dagger(0,0) \rangle$ with $c_{a\sigma}(\textbf{r},t) = \int d^Dq \, e^{i\textbf{q}\cdot\textbf{r}} c_{a\textbf{q}\sigma}(t) $ becomes
\begin{align}
G_\text{lead}&(\textbf{r},t)~\rightarrow ~{\bar G}_\text{lead}(\textbf{r},t) \nonumber \\
& \big\langle e^{-i\frac{\pi^D}{L^D}
\sum_{\textbf{q}} e^{-i\textbf{q}\cdot \textbf{r}}\sqrt{V_\textbf{q}} Q_{\textbf{q}}(t)} 
e^{i\frac{\pi^D}{L^D}\sum_{\textbf{q}} \sqrt{V_\textbf{q}} Q_{\textbf{q}}(t=0)} \big\rangle
\label{Glead}
\end{align}
with ${\bar G}_\text{lead}$ computed with the transformed Hamiltonian. Bohm and Pines show that the renormalization implied by Eq.~(\ref{Glead}) is in essence the RPA reduction in the electronic spectral weight. 

These are exact results. As far as is known, the transformation of the remainder of the lead Hamiltonian can only be carried out approximately, for example, by expanding the exponentials to obtain a series of multiple commutators. Keeping the exact first-order commutator and approximating the second-order term by its vacuum expectation value, $n_0$, Bohm and Pines obtain
\begin{align}\label{H3Db}
H_\text{lead} ~\simeq~ &\sum_{a \textbf{k} \sigma} \epsilon_{\textbf{k}} c^{\dagger}_{a \textbf{k} \sigma} c_{a \textbf{k} \sigma} 
 - \sum_{a \textbf{q}} \sqrt{V_{\textbf{q}}}  Q_{\textbf{q}} \textbf{q} \cdot \textbf{j}_\textbf{q}
 \\ 
&  + \frac{1}{2} \sum_{\textbf{q}} \left[P_{\textbf{q}} P_{-\textbf{q}} +\Omega_p^2(\textbf{q})Q_{\textbf{q}} Q_{-\textbf{q}}\right] 
\nonumber 
\end{align}
with the electron current operator $\textbf{j}_\textbf{q}$ given by
\begin{equation}
i \textbf{q} \cdot \textbf{j}_\textbf{q} = \frac{\pi^D}{L^D} \sum_{\textbf{k}\sigma}(\epsilon_{\textbf{k}+\textbf{q}}-\epsilon_\textbf{k})
c^\dagger_{a\textbf{k}\sigma}c_{a(\textbf{k}+\textbf{q})\sigma} 
\end{equation}
and the plasma frequency $\Omega_p$ defined in terms of the electron stress-energy tensor by
\begin{equation}
\Omega_p^2(\textbf{q})=\frac{\pi^{2D}}{L^{2D}} \, V_\textbf{q} \sum_{\textbf{k}\sigma} n_0  \left(\epsilon_{\textbf{k}+\textbf{q}}+\epsilon_{\textbf{k}-\textbf{q}}-2\epsilon_{\textbf k}\right).
\label{Kdef}
\end{equation}

The coupling between $Q_\textbf{q}$ and the divergence of the fermion current, and the terms dropped in the approximate canonical transformation are to be treated perturbatively. Neglecting them is equivalent to treating the plasmon as an undamped boson and retaining the term in the electron self energy which comes from the electron-plasmon coupling. Adding the $i Q_\textbf{q} (\textbf{q}\cdot \textbf{j}_\textbf{q})$ term in leading order of perturbation theory restores the plasmon damping found within the RPA while the additional neglected terms give beyond-RPA physics.

These transformations have reduced the problem to one analogous to that solved in the previous section.  We now shift the field $P_\textbf{q}$ by $-\sqrt{V_\textbf{q}}n_d$ to remove the dot-lead coupling.  The canonical transformation 
${\cal O}\rightarrow e^{iS}{\cal O}e^{-iS}$ with 
$S=\frac{\pi^D}{L^D} \sum_\textbf{q}\sqrt{V_\textbf{q}}Q_\textbf{q} n_d$ precisely   cancels the Coulomb-induced renormalization of the lead fermion operator, so that in the transformed variables the dot-lead hybridization takes the unrenormalized form
\begin{equation}
H_\text{mix}\rightarrow\sum_{a\textbf{k} \sigma} \left[ \mathcal{T}  d^{\dagger}_\sigma c_{a \textbf{k} \sigma} + \text{h.c.} \right]
\label{Hmix3d1}
\end{equation}
Thus in dimensions greater than $1$ and within the RPA approximation  the Coulombic dot-lead coupling compensates for the Coulomb-induced reduction in electronic spectral weight, so that Coulombic effects drop out of the tunneling problem (except for the 'image charge' or polaron-shift reduction in the dot energy and screening of the dot interaction). The tunneling effects may be calculated using free electrons (that is to say, computing the local lead correlators without including the RPA or GW self energy).

\section{Quantum-dot dynamics}\label{Dynamics}

The previous sections have shown how to reduce the Coulombically coupled dot-lead problem to an  expansion in powers of the dot-lead hybridization which is appropriately renormalized by bosons. The lowest-order term in this expansion implies a master-equation approach to the dot dynamics, which we use to illustrate our formalism. 

For simplicity, we consider a quantum dot with a single (non-degenerate) level described  by a  density matrix  which is diagonal in the occupancy basis and is written as
\begin{equation}\label{rhodiag1}
\rho_d = P_0 |0\rangle \langle 0| + P_1  | 1 \rangle \langle 1|.
\end{equation}
Here $P_0$ and $P_1$ denote the occupation probabilities of the empty state 
$|0\rangle$ and the singly-charged state $|1 \rangle $. Inserting Eq.~(\ref{rhodiag1}) into the von Neumann equation of motion for the density matrix, expanding to leading nontrivial order in the hybridization and using $P_1=1-P_0$ yields the master equation
\begin{align}
\dot{P}_0(t) & = \int_0^{t} dt' \bigg( R_{1 \rightarrow 0}(t-t') \nonumber \\
&-P_0(t') \left[  R_{0 \rightarrow 1}(t-t')+R_{1 \rightarrow 0}(t-t') \right] \bigg) \label{master11} 
\end{align}
with the transition probabilities 
\begin{align}
R_{0 \rightarrow 1}(t) &=2|\mathcal{T}|^2 \, \text{Re} \, \left[F(t) \, e^{-i\varepsilon_d t}\right], \label{R01} \\
R_{1 \rightarrow 0}(t) &= 2|\mathcal{T}|^2 \, \text{Re} \,\left[ F(t) \,  e^{i\varepsilon_d t}\right], 
\end{align}
where $F(t)$ is defined in Eq.~(\ref{Fdef}).

We note in passing that because here we restrict attention to thermal equilibrium,  the long-time limit of the populations is a steady state determined by the detailed-balance condition, i.e.~$P_0$ (as a function of $\varepsilon_d/T$) is a Fermi distribution in the limit $t\to\infty$. Of course, at zero temperature $T=0$, orthogonality effects may cause the system not to equilibrate.

We first examine these equations in the Markovian limit in which the dot dynamics are slow enough and the kernels $F$ decays fast enough that $P_0(t)$ may be treated as a constant and extracted from the integral in Eq.~(\ref{master11}). This yields
\begin{equation}
\dot{P}_0(t) =  \mathcal{R}_{1 \rightarrow 0} - P_0(t) \left(\mathcal{R}_{0 \rightarrow 1}+ \mathcal{R}_{1 \rightarrow 0} \right), \label{rate-equation} 
\end{equation}
where 
\begin{equation}
\mathcal{R}_{0 \rightarrow 1}(\varepsilon_d) = 2|\mathcal{T}|^2 \, \text{Re}\left[ \int_0^\infty d\tau F(\tau) \, e^{-i\varepsilon_d\tau}\right]
\end{equation}
and $\mathcal{R}_{1 \rightarrow 0}(\varepsilon_d)=\mathcal{R}_{0 \rightarrow 1}(-\varepsilon_d)$ are the corresponding Golden-Rule transition rates.

To obtain an idea of the effects of screening we approximate the logarithmic functions $\omega_q$ and $K(q)$ by constants $\omega$ and $K$. The integrals may then be performed analytically and we find  at zero temperature 
\begin{equation}\label{longtime}
F(t) \simeq \alpha \left(\frac{\eta/v}{i t} \right)^{Y}
\end{equation}
for times $t \gg \eta/v$
with
\begin{eqnarray}
Y &=&1-\frac{1}{M}\left(1-\frac{K+K^3}{2}\right),
\\
 \alpha&=&K^{\frac{K+K^3}{2M}},
\end{eqnarray}
if the Coulombic dot-lead interaction is included, and
\begin{eqnarray}
Y& = & 1+\frac{\left(\sqrt{K}-\frac{1}{\sqrt{K}}\right)^2}{2M},
\\
  \alpha&=&K^{\frac{K+K^{-1}}{2M}},
\end{eqnarray}
if not. 

Because for repulsive interactions $0<K<1$ we see that in the presence of Coulombic dot-lead coupling we have $Y<1$, whereas if the coupling is neglected we have $Y>1$. 

\begin{figure}[!t]
\begin{center}
\includegraphics[width=7.8cm,angle=0]{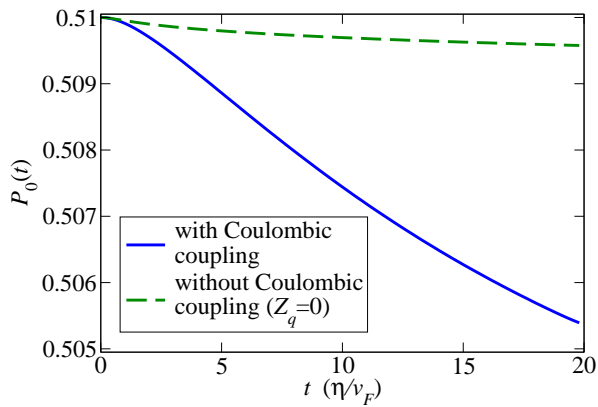}
\caption{(Color online) Time evolution of the occupation probability $P_0(t)$
for the symmetric case, $\varepsilon_d=0$, with initial values $P_0(0)=0.51$ and $P_1(0)=0.49$.
Solid line (blue online): $P_0(t)$ computed from Eq.~(\ref{master11}) with dimensionless coupling $V_c \simeq 0.9$. 
Dashed line (green online): $P_0(t)$ computed from Eq.~(\ref{master11}) but with screening factor $Z_q$ set to zero.
We assume a tunneling amplitude $\mathcal{T}=0.1\, v_F/\eta $ (corresponding to a 
bare tunneling time $\tau_0=16\, \eta/v_F$) and the same parameters as chosen in Fig.~\ref{phi}.}\label{epsilon-dependence} 
\label{timedep}
\end{center}
\end{figure}

The rates are then determined as
\begin{equation}\label{Golden-Rule}
\mathcal{R}_{0 \rightarrow 1}(\varepsilon_d) = \frac{2\pi \mathcal{T}^2}{v /\eta} \,  \left(\frac{|\varepsilon_d|}{v /\eta}\right)^{Y-1} \, \frac{\alpha \ \theta(-\varepsilon_d)}{\Gamma(Y)}.
\end{equation}
Equation (\ref{Golden-Rule}) is of the form of a basic tunneling rate,
\begin{equation}
\frac{1}{\tau_0} = \frac{2\pi {\mathcal T}^2}{v/\eta},
\end{equation}
times a factor expressing the effect of correlations. 
We see that if the Coulombic dot-lead coupling is neglected, the interactions suppress the tunneling rate, whereas in the presence of  Coulombic dot-lead coupling the relaxation rate is enhanced. In the symmetric case $\varepsilon_d=0$ the Markov rate vanishes if the coupling is neglected but diverges if is retained. The steady state of the system obtained from the master equation, Eq.~(\ref{rate-equation}),
depends on the value of $Y$. In the most interesting case, $\varepsilon_d=0$, it is known that although the Markov rate vanishes for $Y>1$,  for $Y<2$ it remains the case that as $t\to\infty$ the occupancy $P_0(t)$ tends to the thermal-equilibrium value, $P_0(t)\to 1/2$, while for $Y>2$ the system does not equilibrate at zero temperature.\cite{Weiss}  In the nanotube case of interest here the effective exponent $K(q)$ vanishes as $q\rightarrow 0$ so that at sufficiently long scales the model without Coulombic dot-lead coupling would fail to equilibrate; however because the increase is only logarithmic,  for reasonable nanotube parameters the effective exponent would only become greater than $2$ for $q \ll \Lambda$.

Figure~\ref{timedep} presents the  time evolution of  $P_0(t)$  obtained by solving the master equation, Eq.~(\ref{master11}), for a dot coupled to a nanotube with and  without the dot-lead Coulombic coupling. We have considered the particle-hole symmetric case $\varepsilon_d=0$ for which the equilibrium value is $P(0)=1/2$ and have begun the simulation in a non-equilibrium initial condition. The strong enhancement of relaxation by the Coulombic dot-lead coupling is evident. 

\section{Conclusions}\label{Conclusions}

In summary, we have studied the effect of a Coulombic dot-lead coupling on the dynamics of a quantum dot. This coupling is always present, but its effects seem not heretofore to have been examined.  We find that it has an important effect on the dot-lead dynamics. Two cases have been considered: a dot coupled to a Luttinger liquid and a dot coupled to two- or three-dimensional metallic leads.  The effects are particularly profound for 
one-dimensional leads. It is well known from previous work that in the absence of the Coulombic dot-lead coupling  the Luttinger liquid correlations of a one-dimensional lead strongly reduce the dot-lead hybridization, leading (at low enough scales and for strong enough interactions) to a complete suppression of tunneling and failure of
the system to equilibrate at $T=0$. The dot-lead Coulomb interaction is shown to overcompensate for this effect, leading to a divergence in the dot-lead hybridization. 

The enhancement of the electronic tunneling due to 
the Coulombic dot-lead interaction has the following 
qualitative interpretation. 
The suppression in the usual case comes because when adding a charge one
has to push the other charges in the Luttinger liquid aside, 
while the presence of the screening interaction 
implies that other charges need not be pushed away
because a screening cloud has to be formed.

We presented estimates for parameters appropriate to a carbon nanotube, which is one of the most widely used one-dimensional leads.  We also presented explicit formulae which could be used for more detailed numerical simulations along 
the lines of Refs.~\onlinecite{Werner07} and \onlinecite{Werner10}. For three-dimensional leads, the effects are found to be less dramatic, but still significant: the consequence of the Coulombic dot-lead coupling is that the Coulombic renormalizations drop out of the problem, so that the dot conductance should be studied using lead Green functions unrenormalized by the RPA or GW corrections to the electron effective mass and scattering. 

Our results rely on several approximations. The most crucial is that density fluctuations in the leads can be represented as noninteracting bosons. The standard results of Luttinger liquid theory \cite{Haldane81} justify this approximation for the case of one-dimensional leads (at least in the universal low-energy limit) while for the case of 
higher-dimensional leads our approximations are at the same level as the random phase approximation. 

Our paper leaves several avenues for future research. Our explicit results are perturbative in the dot-lead hybridization. A numerical or analytical treatment to all orders, leading in particular to an expression for the linear response $I-V$ curve, would be very valuable. An extension of the work to the non-equilibrium case of non-zero bias voltage  is also important. 

\acknowledgments
AJM acknowledges support from the National Science Foundation under grant DMR-0705847 and the hospitality of the Kavli Institute for Theoretical Physics which is supported in part by the National Science Foundation under Grant No.~PHY05-51164. FE acknowledges support from the
Deut\-sche For\-schungs\-ge\-mein\-schaft.

\end{document}